\documentclass[prb,draft,twocolumn,aps]{revtex4}

\usepackage[final]{graphicx}
\usepackage{t1enc}

\begin{document}

\title{\bf Aging dynamics of $\pm J$ Edwards-Anderson spin glasses}

\author{Daniel A. Stariolo}
\email{stariolo@if.ufrgs.br}
\homepage{http://www.if.ufrgs.br/~stariolo}
\affiliation{Departamento de Física, Universidade Federal do Rio Grande do Sul,
CP 15051, 91501-970 Porto Alegre, Brazil}
\author{Marcelo A. Montemurro and Francisco A. Tamarit}
\email{mmontemu@famaf.unc.edu.ar,tamartit@famaf.unc.edu.ar}
\affiliation{Facultad de  Matem\'atica, Astronom\'{\i}a  y F\'{\i}sica, 
Universidad Nacional de C\'ordoba, \\ Ciudad Universitaria, 5000 C\'ordoba,
Argentina}

\date{\today}

\begin{abstract}
We  analyze by  means of  extensive computer  simulations the  out of
equilibrium dynamics  of Edwards-Anderson spin glasses in $d=4$  and
$d=6$   dimensions   with   $\pm J$ interactions.  In
particular, we   focus  our  analysis on   the scaling   properties of
the two-time autocorrelation function in a range of temperatures  from
$T=0.07 T_c$ to $T=0.75 T_c$ in both systems. We observe that  the
aging dynamics of the  $\pm J$ models is  different from that
observed in the  corresponding Gaussian models.  In both the  $4d$ and
$6d$ models  at very low temperatures  we study the effects of discretization
of energy levels. Strong interrupted  aging 
behaviors are found. We argue that
this is because in the times accessible to our simulations the systems are
only able to probe activated dynamics through the lowest discrete energy levels
and remain trapped around nearly flat regions of the energy landscape.
For  temperatures $T
\geq 0.5 T_c$ in $4d$ we find logarithmic scalings that are compatible
with dynamical ultrametricity, while in $6d$ the relaxation can also be
described by super-aging scalings.
\end{abstract}


\maketitle

\section{INTRODUCTION}

After more than  twenty years of  extensive research, the  physics  of
spin  glasses   is  still   far  from   being  completely  understood.
The  inherent  complexity  of  the  physical  scenario  together  with
unsurmountable mathematical difficulties  undermine even the  simplest
theoretical  approaches  that  attempt  to  include  basic   realistic
ingredients. Notwithstanding that a large amount of information  could
be extracted  from the  analysis of  the mean  field model  (i.e.  the
Sherrington-Kirkpatrick  model\cite{mpv,mprrz}),  the   extension   of
those results  to  finite dimensional spin  glasses  remains a  matter
of   hard   controversy    in   the   statistical   physics  community
\cite{droplet,ns}. In  this context large scale  numerical simulations
emerged  as  a valuable  aid  in gaining  physical  insight into  more
realistic models\cite{rieger1,mpr}; however, they  pose such a serious
demand  over existent computational  capabilities that  the study   of
the low temperature phases of these systems may be considered in  many
respects to be still  in an exploratory state.

Regarding  the out  of  equilibrium   behavior of  finite dimensional
spin glasses,  much
attention   has  been  devoted to  models with   continuous couplings
distributions~\cite{mpr,bb}. One important feature of these systems is
that the  ground state  is unique.  For Gaussian  couplings, the aging
behavior of models in $3d$  and $4d$ reported in the  literature seems
to  be compatible  with the  simple aging  or weak  interrupted aging
scenarios (to be defined below). These particular scenarios suggest  a
rather simple phase space structure with a unique relevant time scale,
namely the age of the system. This apparently simple behavior could be
a consequence of the existence  of strongly separated temporal  scales.
Because of this during the time span of a simulation or experiment at
very low temperatures
the system could not be able to cross over between different time
regimes and only one relevant time scale is probed. 

One  may  wonder  whether  the  relatively  simple  dynamical behavior
observed in  spin glasses  with continuous  distributions at  very low
temperatures can also be expected in systems with discrete  couplings,
which present a strong  degeneracy of their ground states as  well  as
a noticeable discretization of their low-energy spectrum. That   might
lead to very low temperature  effects that could not be observable  in
continuous spin glasses.

It  is known   that at  temperatures  relatively close to  $T_c$ there
is no qualitative difference in the phenomenological  characterization
of aging \cite{rieger2} between the continuous and discrete couplings
systems. However,  at lower   temperatures, effects  deriving from the
discrete  nature of  the energy  spectrum as  well as  from the  high
degeneracy  of  the  ground states  may  become  apparent in  discrete
models.

Scaling properties  contribute to a  quantitative description  of complex
phenomena,  even  in   cases  where  a   general  theory  is   lacking
\cite{droplet,bckm,yohuta02}. In this  respect,  different  dynamical  universality
classes  may  emerge for which similar scaling functions  describe the
dynamics of different  systems. The knowledge  of these scaling  rules
gives  considerable  insight  into  the  nature   of  the   underlying
dynamical processes.

In this  paper we present the results of an extensive  numerical study
of  the  aging dynamics  and  scaling  properties   of   the  two-time
autocorrelation  functions    for the  $\pm  J$ Edwards-Anderson  spin
glasses in dimensions $d=4$ and $d=6$ at  temperatures which cover the
whole low temperature phases of the models.

Our results show important  differences between the aging  dynamics of
the $\pm J$  models compared with  the corresponding Gaussian  models.
At the very low temperatures  $T=0.07 T_c$ and  $T=0.15 T_c$ we  find
interrupted aging scenarios both in $4d$ and $6d$. In particular,  for
the lowest temperature we find  a very small exponent indicative  of a
rapid relaxation to a stationary  dynamics. At first sight this  seems
hard to reconcile with the  expected very slow relaxation which  takes
place at these extremely low temperatures. Nevertheless this can be seen as a
consequence of the discrete nature of energy levels which can produce
some non trivial time dependent phenomena at very low temperatures
at time scales roughly independent of the waiting time $t_w$. The systems
simply do not have enough time to relax over barriers
dependent on $t_w$. So this dynamics is completely different from the
usual long time non-equilibrium relaxation but is not trivial due to
the discreteness of the low lying energy levels.
At $T=0.5 T_c$ we observe some differences in the  behavior
in $4d$ and in  $6d$. In $4d$ logarithmic  scalings, which  are compatible
  with dynamic  ultrametricity,  work very well  even at this relatively
high temperature. 
In $6d$ the scenario is less clear since we  observe
that although logarithmic scalings are very good it is also
possible to scale the data with a {\em super-aging} scaling form. As $d=6$
is the upper critical dimension the relaxation dynamics should present
some features of the infinite range model. We find insted that the behavior
is more similar to that at lower dimensions than to the mean field model.

The paper is organized as follows:  in section II we define the  model
studied and the observables measured; in section III and IV we analyze
the  results  of  the  4d  and  6d  models respectively.  Finally the
conclusions and a discussion are presented in section V.

\section{MODEL AND METHOD}
The system  consists of  a d-dimensional  hypercubic lattice  of Ising
spins which interact according to the following Hamiltonian:
\begin{equation}
{\mathcal{H}}= -\sum_{\langle i,j \rangle} J_{ij} S_i S_j \; ,
\end{equation}
where  the  symbol $\langle  i,j  \rangle$ indicates  that  only first
neighbor pairs $i,j$ are  taken into account. The coupling   constants
$J_{ij}$  are  binary random   variables chosen   from  the  following
probability distribution:
\begin{equation}
\rho(J_{ij})= \frac{1}{2}\Bigl( \delta(J_{ij} -1) + 
\delta(J_{ij}+1)\Bigr) \; .
\end{equation}
The time evolution  of the model  is governed by  a standard heat-bath
Monte  Carlo  process   with sequential  random   update.  The imposed
boundary conditions were periodic  in 4d and helical  in 6d~\cite{mc}.
In  the  practical  implementation  of   the  numerical   algorithm  a
significant increase in speed  was  accomplished by  using  multi-spin
coding \cite{mc} so  each spin  and coupling constant demand just  one
bit of  information each  for storage.  In this  way  we  can run many
replicas at the same  time at the cost   of a single realization.   In
all cases  the  dynamics is  initiated  from a random   configuration,
simulating   a  sudden  quench    from  infinite temperature into  the
spin glass phase.

One  straightforward  way  to  characterize  the  out  of  equilibrium
dynamics of complex magnetic  systems is through  the analysis of  the
two-time  autocorrelation  function   $C(t,t')$,  which   can  exhibit
history dependent features   usually referred  to as  {\em aging}.  A
system   that  has  attained  thermodynamic  equilibrium  will  show a
stationary dynamics for which  only  time  differences  make  physical
sense,   and   therefore $C(t,t')\equiv   C(t-t')$.  However,  complex
magnetic  systems  such  as  spin glasses  show a  much  more  complex
behavior  due  to  the  presence  of  an  extremely  slow relaxational
dynamics. These materials may be out of  equilibrium for spans of time
longer  than any  available time  scale in   the laboratory.  In this
circumstances insight  into  the  ongoing  processes  can  be obtained
by  studying  the  scaling  properties  of  dynamical  quantities like
$C(t_w+t,t_w)$, where the  {\em waiting time}, $t_w$, stands for   the
{\em age} of the  system measured  after a quench into the spin  glass
phase, and $t$ stands for the  time measured since the age $t_w$.

In the numerical experiments we compute the quantity
\begin{equation}
C(t_w+t,t_w)=\Bigl [ \frac{1}{N}\sum_{i=1}^N S_i(t_w+t)S_i(t_w)
\Bigr ]_{av} \ ,
\label{web}
\end{equation}
where we have denoted by $[\ldots]_{av}$ an average taken over several
realizations   of the   random couplings  and thermal  histories.
An additive form for the autocorrelations was assumed:
\begin{equation}
C(t_w+t,t_w)=C_{st}(t) + C_{ag}\Bigl( \frac{h(t_w+t)}{h(t_w)} \Bigr)
\ .
\label{additive}
\end{equation}
The  stationary part  $C_{st}(t)$ is  well described  by an  algebraic
decay of  the form $C_{st}(t)=At^{-x(T)}+q$. 
 There is no theoretical basis for determining  the
scaling  function  $h(z)$   appearing  in  the   aging  part  of   the
autocorrelations. There have  been proposed a  couple of simple  forms
which work reasonably well for  relaxations in the vicinity of  simple
aging, in the sense that the  correct scaling variable is of the  form
$t/t_w^{\mu}$ with  $\mu \sim  1$ \cite{bb}.  In our  case it  was not
possible to find  a unique realization  of the function  $h(z)$ for
the whole range  of temperatures studied.  In the scaling  analysis we
restricted ourselves to  find the relevant  scaling variables for  the
aging part of the autocorrelations.

\section{d=4}

The simulations of the four dimensional Edwards-Anderson model were
done for systems of linear size $L=12$ $(T=0.07T_c)$ and $L=10$ imposing 
periodic boundary
conditions. We recall that the critical temperature for this model 
has been estimated to be $T_c \approx 2$\cite{hu99,mazu99}.

\subsection{Very low temperatures: probing the discreteness of the lowest 
energy levels}

\begin{figure}[ht]
\includegraphics[width=7cm,height=8.5cm,angle=270]{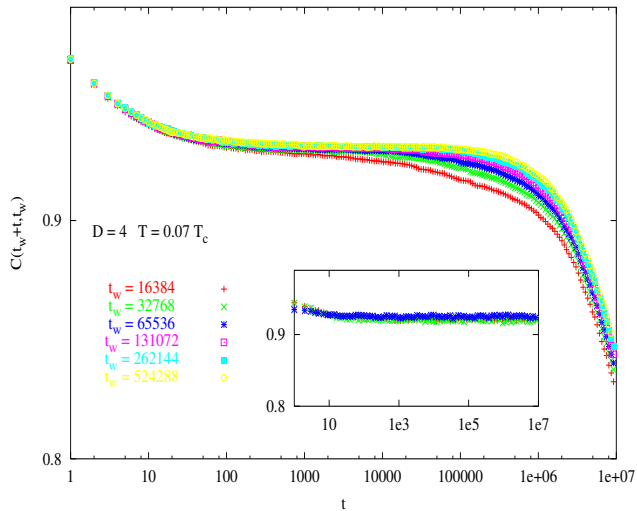}
\caption{\label{figure1}
Two-time  autocorrelation  function  for the  $\pm  J$  spin glass  in
$d=4$ with linear size $L=12$ after a quench to $T=0.07 T_c$ in a double logarithmic plot.
The waiting times are $t_w=2^k, k=14\ldots 19$ from bottom to top. Inset: the
result of a microcanonical run after $t_w$ for the three longest tw' s.}
\end{figure}

In   Figure~\ref{figure1}  we  show   the  behavior  of   the two-time
autocorrelation  function  $C(t_w+t,t_w)$  at the temperature  $T=0.07
T_c$.  The  waiting  times  are  $t_w=2^k,  k=14\ldots  19$,  and  the
simulation  was   run  up to $t=10^7$  MCS. There is a weak dependence
on the wainting time and the system relaxes to a stationary regime
presenting a strong sub-aging behavior.

\begin{figure}[ht]
\includegraphics[width=7cm,height=8.5cm,angle=270]{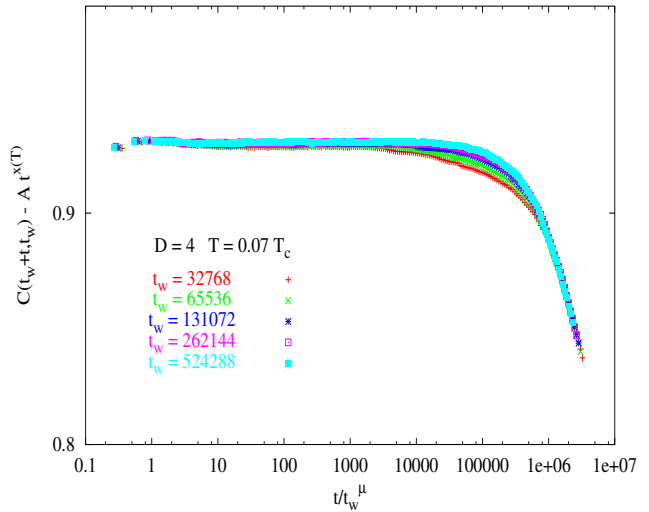}
\caption{\label{figure2}
Interrupted aging scaling for the asymptotic time regime of the data 
shown in figure \ref{figure1}.
The stationary decay of the correlations has been subtracted, fit
parameters are $A=0.04$ and $x=0.57$. The interrupted aging
exponent is $\mu=0.1$.}
\end{figure}

In Figure~\ref{figure2} we show the best scaling obtained for the long
time behavior of the autocorelation function after subtraction of  the
stationary part:
\begin{equation}
C_{st}(T)= At^{-x(T)} + q.
\label{stationary}
\end{equation}

In this particular case, the best fit was obtained for $x(T)=0.57$  (a
rather high value)  and $q \approx 0.93$. In the (weakly) $t_w$ dependent
region it is possible to distinguish two regimes: a  first one  when the  system
falls  out   of  the   quasi-equilibrium  and   a  second   regime  at
long  times. Both are very  well  described   by different
interrupted or sub-aging scalings of the form:
\begin{equation}
C_{ag}(t_w+t,t_w)= {\mathcal C}\Bigl(\frac{t}{t_w^{\mu}} \Bigr) \ .
\label{interrupted}
\end{equation}

\begin{figure}[ht]
\includegraphics[width=7cm,height=8.5cm,angle=270]{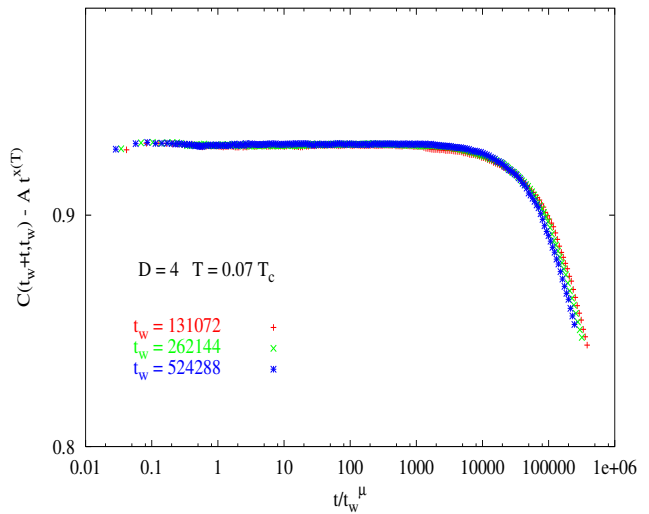}
\caption{\label{figure3}
Interrupted aging scaling for the intermediate time regime of the data 
shown in
figure \ref{figure1}. This time regime is characterized by an
interrupted aging exponent $\mu=0.27$
higher than the corresponding one seen in the asymptotic regime of figure
\ref{figure2}.}
\end{figure}

For very long times, the best fit yields an exponent  $\mu \simeq 0.1$  
which characterizes a  very strong sub-aging regime.
It  is  also  possible  to  collapse  the  data  in  the  time  window
corresponding to  the first  out of  equilibrium regime ($t \approx t_w$)
 with the form
given by Eq.(\ref{interrupted}) with an exponent $\mu \simeq 0.27$  as
can  be  seen  in  figure \ref{figure3}.  Note that the first  sub-aging 
regime  disappears gradually  as
$t_w$ grows  and effectively,  after the  longest waiting  time studied
$t_w  \simeq  5\cdot  10^5$,  these  slow  modes   have  practically
equilibrated. After these very  long waiting times the  system evolves
quickly to  a stationary  situation with  a very  short characteristic
aging  time   $\simeq  t_w^{0.1}$.
The observation of these sub-aging
regimes at this  very low  temperature is consequence of the discrete
nature of the low lying energy levels. In fact the smallest time scale for
activation in this energy landscape is of the order of 
$\exp{(2/T)} \approx 1,6\cdot 10^6$ which is approximately the time where the
correlation leaves the quasi-equilibrium regime (see Figure~\ref{figure1}).
Up to this time scale the system relaxes effectively in the flat energy
landscape defined by the sites with zero local field which are free to
flip (besides a smaller group which can still contribute to lowering the
energy). Note that the value of the correlation in the plateau $C=q$ in this
regime does not correspond to the equilibrim order parameter $q_{EA}$ but
instead to one minus the fraction of sites with zero field
\footnote{This was pointed to us by Federico Ricci-Tersenghi (private communication).}.
 After a time of the order of $\exp{(1/0.07)}$ thermal activation begins
to take place through the lowest lying barriers and the correlations
decay to zero as the system expands its available phase space. As a further
test for this interpretation we have performed a set of simulations in a
microcanonical ensemble which are shown in the inset of Figure~\ref{figure1}:
for the three longest waiting times we have allowed the system to relax up
to $t_w$ and from then on only flips of spins with zero local fields were allowed.
In these conditions the correlations decayed only to the plateau suggesting
that the long time decay in the canonical dynamics was produced by activation 
over low energy barriers. The scenario for this behavior is quite clear:
after a long waiting time, the system diffuses further in a flat energy
landscape surrounded by barriers of minimum height $\Delta E=2$. At time
scales of the order of $\exp{(2/T)}$ it can further relax by thermal
activation over these low barriers. The next level is at $\tau \simeq 
\exp{(4/T)} \approx 2,56\cdot 10^{12}$ and it is clearly unreachable. The
simplicity of the landscape seen by the system at this temperature explains
the presence of the rapidly interrupted aging which is observed.

\begin{figure}[ht]
\includegraphics[width=7cm,height=8.5cm,angle=270]{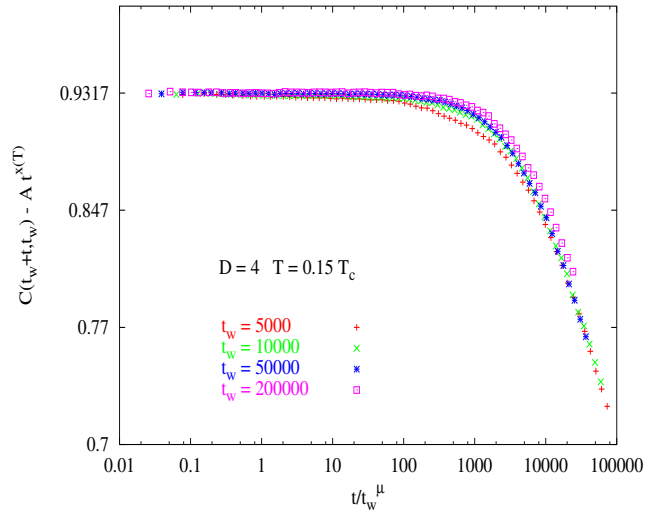}
\caption{\label{figure4}
Interrupted aging scaling for the asymptotic regime of the d=4
Edwards-Anderson 
model at $T=0.15 T_c$ with linear size $L=10$ and four different waiting times
$t_w=5000, 10000, 50000$
and $200000$. The fitting parameters of the stationary regime are
$A=0.03$ and $x(T)=0.54$. In this regime the sub-aging exponent is 
$\mu=0.3$. Note that the curve for the longest waiting time is beginning
to depart from this regime.}
\end{figure}

In figure \ref{figure4} we present  the
results of a simulation performed at a higher temperature  (though
still very low) $T=0.15 T_c$. At this temperature in the time range of the 
simulation the system is able to probe activation over barriers of height
$\Delta E=2$ and $\Delta E=4$. The waiting times are $t_w=5000$, $10000$, $50000$
and  $200000$.  The  three  curves for the shortest waiting  times  
can still be  well
collapsed    for  large  values  of  $t$  by  the   interrupted aging
 form (\ref{interrupted}), with an exponent $\mu\approx 0.3$.  However,
the last  curve ($t_w=200000$)  clearly departs  from this  scaling,
signaling a  crossover to  another regime. It is  still
possible to see an intermediate regime for the shortest $t_w$'s.

\subsection{Full aging dynamics}

\begin{figure}[ht]
\includegraphics[width=7cm,height=8.5cm,angle=270]{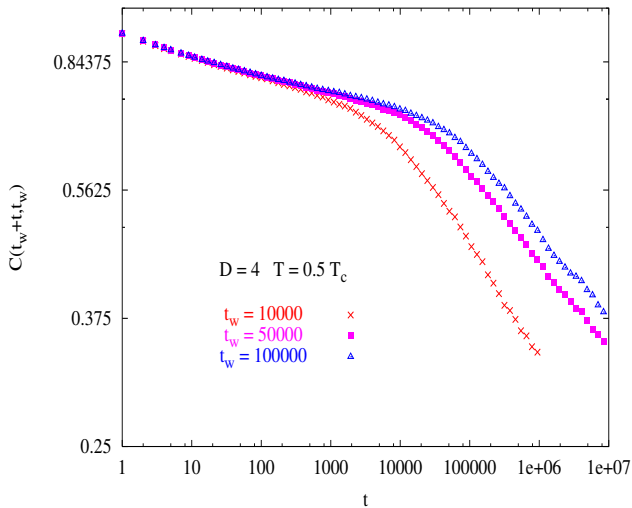}
\caption{\label{figure5}
Two-time autocorrelation function for the $\pm J$ spin glass
in $d=4$ with linear size $L=10$ after a quench to $T=0.5 T_c$.
The waiting times are $t_w=10000, 50000$ and $100000$.}
\end{figure}

\begin{figure}[ht]
\includegraphics[width=7cm,height=8.5cm,angle=270]{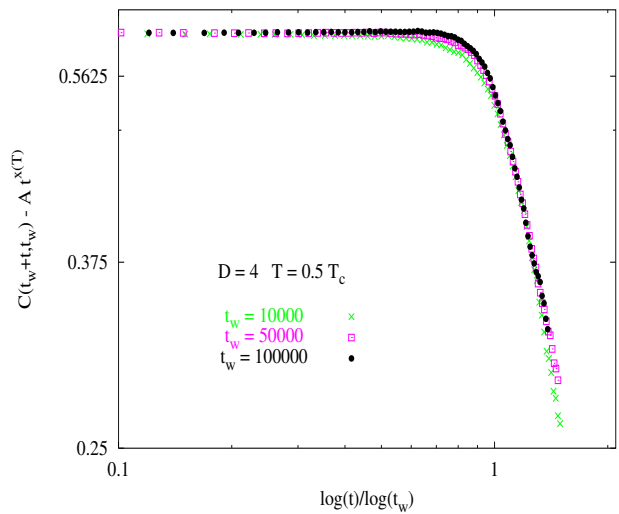}
\caption{\label{figure6}
Logarithmic aging scaling for the asymptotic time regime of the data 
shown in figure \ref{figure5}. Fit parameters are $A=0.3$ and $x(T)=0.1$.}
\end{figure}

As the thermal energy is raised above the lowest lying levels the full
ruggedness of the landscape should emerge and true aging dynamics should
be restored.
What happens when the thermal energy is enough to turn discrete energy
levels   undetectable?   In   figure   \ref{figure5}   we   see    the
autocorrelations for a temperature $T=0.5 T_c$ for three waiting times
$t_w=10000, 50000, 100000$.  A  fit  to  the  quasi-equilibrium  region  gives
$x(T)=0.1$  and  $q=0.62$, indicating  a very slow relaxation.
This is due to the increasing  complexity of  the phase  space visited
at this temperature. The best scaling form obtained for  the
aging regime is presented  in figure \ref{figure6} and  corresponds to
the following logarithmic form:
\begin{equation}
C_{ag}(t_w+t,t_w)= {\mathcal C}\Bigl(\frac{\ln{t}}{\ln{t_w}} \Bigr)\ .
\label{ultrametric}
\end{equation}
Note that  there is  still an  intermediate regime  which is  not well
described   by this  logarithmic scaling  but the  corresponding time
window is too small to try a reasonable collapse there. On the other side, in
the long time regime  the  logarithmic  scaling (\ref{ultrametric})
works  remarkably well. It is worth noting that this form of the
scaling  function is   compatible with  dynamical ultrametricity  (see
\cite{bertin}). This form is slightly different from that expected  by
a droplet like scenario,  which takes the form  $\ln{(t+t_w)}/\ln{(t_w)}$
and does not obey ultrametricity. This new evidence for a (weak)  ultrametricity
obtained directly from aging measurements is in agreement with  recent
results for the  same model obtained  from a quite  different approach
\cite{st01}.  
We have also done simulations at temperature $T=0.75 T_c$ (not shown) and the 
results are
compatible with the logarithmic scaling (\ref{ultrametric}) although due to
the strong noise at this rather high temperature the data is not so clean
as at lower temperatures and the scaling of the curves is not so good. At
high temperatures several time scales are mixed or superposed
because of thermal noise and consequently it is difficult to obtain
reliable scalings. 

\subsection{d=6}
We have done a similar analysis of data obtained
for  the  $d=6$  Edwards-Anderson model. As this model is at the upper
critical dimension we expected to see a behavior similar to the
mean field or Sherrington-Kirkpatrick (SK) model 
\cite{cuku94,maparo98,motastca00}.
The off equilibrium dynamics of the SK model is only poorly understood due
to the complexity which emerges from the full replica symmerty breaking
which is the central characteristic of that model. As a consequence of the
complex ultrametric organization of time scales no simple scaling form can
be found for the aging dynamics which might be described by a superposition
of scaling regimes of the form ~\cite{cuku94}:

\begin{equation}
C_{ag}(t_w+t,t_w)= \sum_i {\mathcal C_i} \Bigl(
\frac{h_i(t_w+t)}{h_i(t_w)} \Bigr) \ .
\label{suma}
\end{equation}

Nevertheless
we have found a much simpler scenario which is in fact very similar to what
is observed in $d=3$ and $d=4$. The transition to an SK like aging scenario
seems to be very slow as the connectivity of the system grows. 

 All  these
simulations were carried out for systems with linear size $L=5$. It is
worth noting that the computational time required for simulating  these
systems increases  as $L^6$,  making it  much more  difficult  to  use
larger values of $L$. The critical temperature of this model was estimated
in \cite{wayo93,be96} to be $T_c \simeq 3$. Here again we start by 
considering the very  low temperature case $T=0.07 T_c$. 

\begin{figure}[ht]
\includegraphics[width=7cm,height=8.5cm,angle=270]{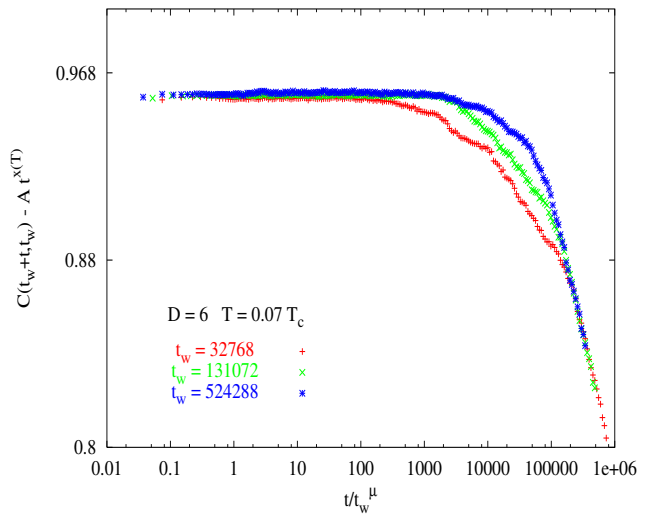}
\caption{\label{figure7}
Sub-aging scaling for two-time autocorrelation
functions of the d=6 Edwards-Anderson $\pm J$ spin glass
 at $T=0.07T_c$. The fit parameters are $A=0.027$, $x(T)=0.55$
and $\mu=0.25$.}
\end{figure}

\begin{figure}[ht]
\includegraphics[width=7cm,height=8.5cm,angle=270]{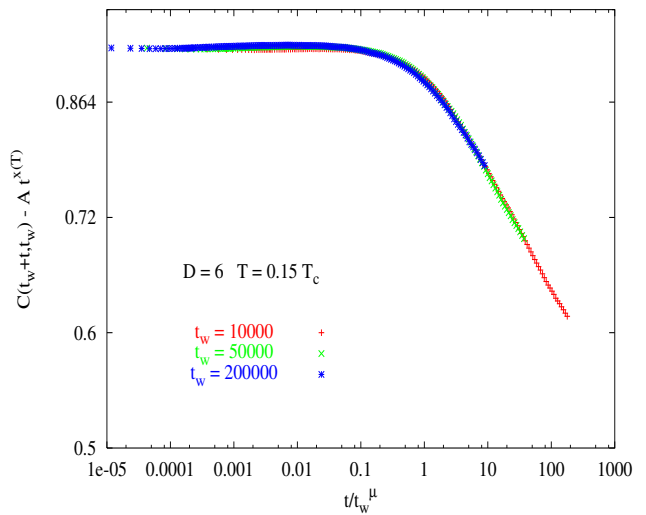}
\caption{\label{figure8}
Sub-aging scaling for the two-time autocorrelation
functions for the d=6 $\pm J$ spin glass at $T=0.15T_c$. 
Fit parameters are $A=0.04$, $x(T)=0.29$ and $\mu=0.93$.}
\end{figure}

The  overall behavior  of the  two-time autocorrelation  function is
similar  to that  described for  dimension $d=4$.  It is  possible to
identify two different sub-aging regimes within the time window of  our
simulation, which covers  up to $t=10^7$.  Both  in   the
intermediate   and   in   the   asymptotic   regimes   the   functions
$C(t_w+t,t_w)$ can be very well collapsed by an interrupted aging scenario of
the form (\ref{interrupted}). In Figure~\ref{figure7} we show the best
scaling  obtained for  the long  time behavior  of the  autocorrelation
after subtraction of the stationary part. The plots correspond to  the
three longest waiting times: $t_w=2^{15}$, $2^{17}$ and $2^{19}$.  The
fitting  in  the  quasi-equilibrium  regime  yielded  $x(T)=0.55$  and
$q \simeq 0.96$,  similar  to  those   found  for  $d=4$  at   the  same
temperature. The two regimes described above can be clearly identified
in this graph. For the last  regime, the best data collapse yields  an
exponent  $\mu  \approx  0.25$ and  in  the  intermediate regime  $\mu
\approx 0.9$ (this figure is not shown). 

In  figure  ~\ref{figure8}  we  present  the  results  of  a   similar
simulation performed at  $T=0.15 T_c$. Here  we note that the long times
regime which had a smaller exponent (faster relaxation) has now disappeared. 
Actually the three curves can be very well
collapsed by a unique  interrupted scaling form with  $\mu=0.93$. The
main difference between this  simulation and that performed  for $d=4$
(see figure \ref{figure4}) is the  fact that now it is not seen
the crossover described in  $d=4$ for the longest waiting time.

Note also that as $T$ increases  the scaling  
form  approaches the simple aging scaling law $\mu \to 1$. 

\begin{figure}[ht]
\includegraphics[width=7cm,height=8.5cm,angle=270]{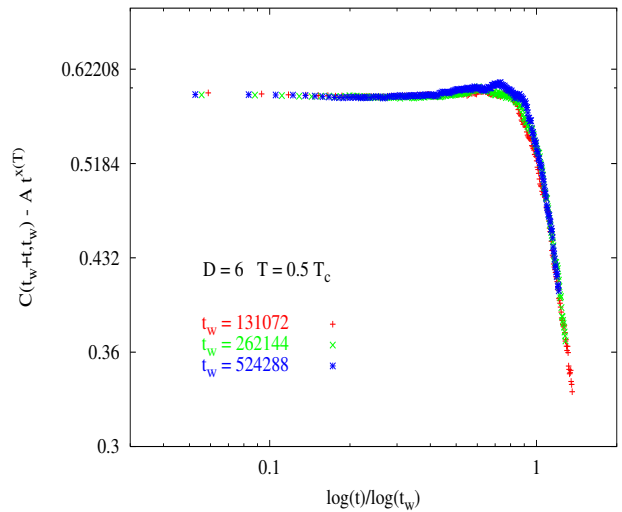}
\caption{\label{figure9}
Logarithmic  scaling  for  the two-time
autocorrelation functions of the d=6 Edwards-Anderson spin glass at $T=0.5T_c$.
Fit parameters are $A=0.34$ and $x(T)=0.16$ and $q=0.59$.}
\end{figure}

\begin{figure}[ht]
\includegraphics[width=7cm,height=8.5cm,angle=270]{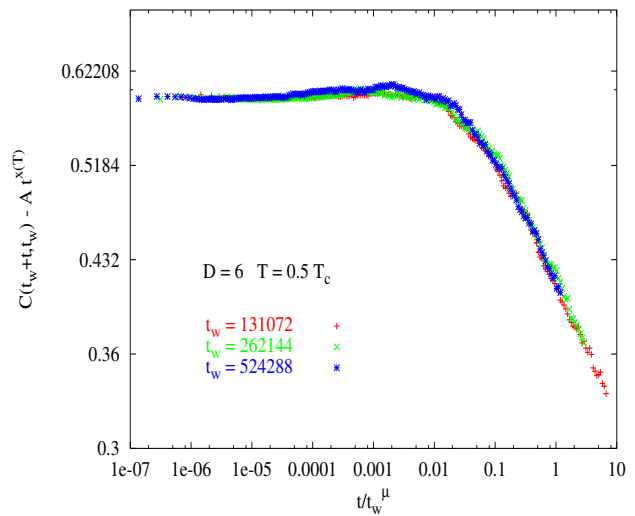}
\caption{\label{figure10}
Super-aging  scaling  for  the same data of figure (\ref{figure9}).
Fit parameters are $A=0.34$ and $x(T)=0.16$, $q=0.59$
and $\mu=1.2$.}
\end{figure}

At  $T=0.5T_c$  the  scenario  is  not  so  clear  as  in $d=4$.
In figures \ref{figure9} and \ref{figure10} we  present
two  different  scalings.  In  \ref{figure9}  we  use  the expression
(\ref{ultrametric}) and in \ref{figure10}  the same scaling form  used
in the interrupted scenario.  Nevertheless, it is important  to stress
that in this last case the exponent that produces the best collapse is
$\mu\approx 1.2$, indicating  the possible existence  of a super-aging
regime. However,  as can  be concluded  from these  figures, it is difficult
to decide  which is the  true scaling law.
In summary, as mentioned at the beginning of this section the $d=6$ model
presents an aging dynamics very similar to the $d=4$ case at corresponding
temperatures and it seems to be still far from the limit of the mean field 
system studied in \cite{maparo98,motastca00}.

\section{DISCUSSION AND CONCLUSION}

In this paper  we have presented  an extensive numerical  study of the
out  of  equilibrium  dynamics of spin  glasses with  discrete couplings
  defined  on
hypercubic  lattices  in  $4$  and  $6$  dimensions,  for very low 
temperatures $T=0.07T_c$, $T=0.15T_c$ and $T=0.5T_c$. 

The relaxation behavior of the model with $\pm J$ couplings is different
from that with continuous gaussian couplings.
While the observed aging dynamics of the gaussian model is described
in a wide range of temperatures and time scales by simple aging or
weak interrupted aging scalings, the observed behavior of the $\pm J$
model seems to depart from that. 
At the lowest temperatures studied , that is for $T=0.07T_c$ and $T=0.15T_c$,
the thermal energy is still not enough to permit activation over several
scales and only the lowest lying energy levels are probed. As a consequence
the system diffuses on a flat energy landscape surrounded by low barriers
of height 2 or at most 4. It relaxes in this simple landscape presenting
a kind of interrupted aging with a rapid relaxation to stationarity. In this
sense these relaxations are practically independent of $t_w$ and true 
aging is restored only at higher temperatures. In order to see true aging
at these very low temperatures we would have to wait for extremely long times.
 At $T=0.5 T_c$ both in $d=4$ and $d=6$ the full aging is restored, no more signs of
interrupted scalings are seen but a slower logarithmic scaling is observed.
At this intermediate temperature range thermal energy is large compared with
the low lying energy states which were important at the lower temperatures.
Consequently the aging dynamics proceeds slowly but with the tendency of
restoring ergodicity and equilibrium as $t_w$ grows.
 It is important to stress that for no one of the temperatures
studied $T=0.07 T_c$, $T=0.15 T_c$, $T=0.5 T_c$, and $T=0.75 T_c$ simple
aging scenarios were found, in contrast to what seems to be the rule for
the gaussian spin glass model. Insted the logarithmic aging observed at
$T=0.5 T_c$ (and also at $T=0.75 T_c$) points to a simple hierarchy of time scales with dynamical
ultrametricity as observed in \cite{st01}. On the other hand, even for the
model in $d=6$, which is at the upper critical dimension, the qualitative
form of the aging curves and the scaling functions are very different from
those found in the SK model\cite{maparo98} and in the Hopfield model
\cite{motastca00} which are known to present full replica symmetry breaking
and a consequently full hierarchical or ultrametric organization of time
scales\cite{cuku94}. In this respect it is worth citing recent work by
Yoshino, Hukushima and Takayama~\cite{yohuta02} where they presented an 
extended version of
the dynamical droplet theory. Two interesting new predicitions are the presence of
a new dynamical order parameter $q_D < q_{EA}$ which should be observed in
a particular lenght/time scaling regime and the distinction between the times at
which time translation invariance and the Fluctuation-Dissipation Theorem
are violated. The authors presented numerical results on the $d=4$ $\pm J$ spin
glass to support their findings but clearly more precise measurements are needed in
order to test the theoretical predictions. In particular, scaling functions depend
on the growth law of the coherence lenght $L(t)$ which permits to see the
crossover between critical to activated dynamics. Unfortunately this information
is still not available for the temperatures and time scales reached in our 
simulations which
we think are necessary in order to clearly see the separation of time regimes
during aging dynamics. We expect new interesting results to come from these kind
of analysis in the near future. 

\begin{acknowledgments}
We thank F. Ricci-Tersenghi for interesting comments on an earlier version of the
manuscript. 
The work of D.A.S. was supported in part by Conselho Nacional de Desenvolvimento
Científico e Tecnológico (CNPq) and Funda\c cão de Amparo à Pesquisa do Estado
do Rio Grande do Sul (FAPERGS), Brazil. M.A.M. acknowledges finacial support
from the Secretary of Science and Technology (SeCYT)
of the National University of Córdoba, Argentina.
\end{acknowledgments}


\begin{thebibliography}{99}
\bibitem{mpv}M. Mézard, G. Parisi and M. Virasoro,
{\em Spin Glass Theory and Beyond}, (World Scientific, Singapore, 1986).
\bibitem{mprrz}E. Marinari, G. Parisi, F. Ricci-Tersenghi, J. J.
Ruiz-Lorenzo
and F. Zuliani, J. Stat. Phys. {\bf 98}, 973 (2000).
\bibitem{droplet}D. S. Fisher and D. A. Huse, Phys. Rev. B {\bf 38}, 373
(1988) and Phys. Rev. B {\bf 38}, 386 (1988).
\bibitem{ns}C. M. Newman and D. L. Stein, Phys. Rev. Lett. {\bf 76}, 515
(1996), Phys. Rev. E {\bf 57}, 1356
(1998) and  preprint cond-mat/0105282.
\bibitem{rieger1}H. Rieger, in {\em Annual Reviews of Computational
Physics},
vol. II, D. Stauffer editor, (World Scientific, Singapore, 1995).
\bibitem{mpr}E. Marinari, G. Parisi and J. J. Ruiz-Lorenzo, in {\em Spin
Glasses
and Random Fields}, A. P. Young editor, (World Scientific, Singapore,
1998).
\bibitem{bb}L. Berthier and J-P Bouchaud, preprint cond-mat/0202069.
\bibitem{hu99}K. Hukushima, Phys. Rev. E {\bf 60}, 3606 (1999).
\bibitem{mazu99}E. Marinari and F. Zulinani, J. Phys. A {\bf 32},
7447 (1999).
\bibitem{rieger2}H. Rieger, {\em J. Phys.} {\bf A26}, L615 (1993).
\bibitem{bckm}J. P. Bouchaud, L. F. Cugliandolo, J. Kurchan and M.
M\'ezard, in {\em Spin Glasses and Random Fields}, A. P. Young editor (World
Scientific, Singapore, 1998).
\bibitem{yohuta02}H. Yoshino, K. Hukushima and H. Takayama, preprint
cond-mat/0202110 and preprint cond-mat/0203267.
\bibitem{mc}M. E. J. Newman and G. T. Barkema, {\em Monte Carlo Methods in
Statistical Physics}, (Oxford University Press, Oxford, 1999).
\bibitem{cuku94}L. F. Cugliandolo and J. Kurchan, J. Phys. A {\bf 27},
5749 (1994).
\bibitem{bertin} E. Bertin and J.-P. Bouchaud, {\em Dynamical
ultrametricity in the critical trap model}, preprint cond-mat/0112187.
\bibitem{st01}D. A. Stariolo, Europhys. Lett. {\bf 55}, 726 (2001).
\bibitem{maparo98}E. Marinari, G. Parisi and D. Rossetti, Eur. Phys. J. B
{\bf 2}, 495 (1998).
\bibitem{motastca00}M. A. Montemurro, F. A. Tamarit, D. A. Stariolo and S. A.
Cannas, Phys. Rev. E {\bf 62}, 5721 (2000).
\bibitem{wayo93}J. Wang and A. P. Young, J. Phys. A {\bf 26}, 1067 (1993).
\bibitem{be96}L. Bernardi, PhD. Thesis, Université di Paris-Sud, Centre 
D'Orsay, 1996.
\end{thebibliography}
\end{document}